\documentclass[final]{article}

\usepackage{amsmath}
\usepackage{amsfonts}
\usepackage{amsthm}
\usepackage{amssymb}
\usepackage{graphicx}
\usepackage{bm}
\usepackage{mathtools}
\usepackage{booktabs}
\usepackage[colorlinks=true,plainpages=false]{hyperref}
\hypersetup{urlcolor=blue, citecolor=red}
\allowdisplaybreaks

\newtheorem*{theorem*}{Local Center Manifold Theorem}

\newcommand{\ii}{\mathrm{i}}
\newcommand\numberthis{\addtocounter{equation}{1}\tag{\theequation}}

\DeclarePairedDelimiter\abs{\lvert}{\rvert}%
\DeclarePairedDelimiter\norm{\lVert}{\rVert}%
\makeatletter
\let\oldabs\abs
\def\abs{\@ifstar{\oldabs}{\oldabs*}}
\let\oldnorm\norm
\def\norm{\@ifstar{\oldnorm}{\oldnorm*}}
\makeatother

\begin{document}

\title{Informing the structure of complex Hadamard matrix spaces using a flow}

\author{Francis C. Motta\thanks{Department of Mathematics, Duke University, Box 90320, Durham, NC 27708-0320. motta@math.duke.edu} \; and Patrick D. Shipman\thanks{Department of Mathematics, Colorado State University, 1874 Campus Delivery, Fort Collins, CO 80523-1874. shipman@math.colostate.edu}}

\date{}

\maketitle


\begin{abstract}
The defect of a complex Hadamard matrix $H$ is an upper bound for the dimension of a continuous Hadamard orbit stemming from $H$. We provide a new interpretation of the defect as the dimension of the center subspace of a gradient flow and apply the Center Manifold Theorem of dynamical systems theory to study local structure in spaces of complex Hadamard matrices. Through examples, we provide several applications of our methodology including the construction of affine families of Hadamard matrices.
\end{abstract}

\section{Introduction}
The principal objects of interest in this paper are \textbf{dephased complex Hadamard matrices}: matrices $H \in M^{d \times d}(S^1)$ such that
\begin{align*}H H^* & =  dI_d, \text{ and } \\
[H]_{i,1} & =   [H]_{1,i} = 1  \text{ for } i=1,\ldots,d.
\end{align*}
Here $I_d$ is the $d \times d$ identity matrix, $^*$ denotes conjugate transpose, and $S^1 \subset \mathbb{C}$ is the complex unit circle.  We will adhere to the common practice of declaring two Hadamards $H$ and $K$ equivalent if there exist unitary diagonal matrices $D_1$ and $D_2$ and permutation matrices $P_1$ and $P_2$ such that $H = D_1P_1 K P_2D_2.$  

Complex Hadamards are natural generalizations of their real analogue: square matrices with entries in $\{-1,1\}$ with mutually orthogonal rows and columns, discovered to have the largest determinant among all real matrices whose entries have absolute values bounded by one~\cite{cite:hadamard}.  While both real and complex Hadamards are mathematically interesting objects in their own right, they also have a wide range of applications including uses in coding theory~\cite{cite:hadamardapp1}, the design of statistical experiments~\cite{cite:hadamardapp2}, numerous constructions in theoretical physics~\cite{cite:zeilinger1,cite:zeilinger2,cite:hadamardapp3} and quantum information theory~\cite{cite:werner}.

The focus of much of the current mathematical research is aimed at complete classification of equivalence classes of complex Hadamards, at least for small dimensions~\cite{cite:knownfamilies2,cite:knownfamilies3,cite:knownfamilies4}.  In dimensions $d \leq 5$, complex Hadamards have been completely described~\cite{cite:4x4hadams,cite:completeclass}, while the classification of $6 \times 6$ Hadamards remains open if not on the precipice of completion~\cite{cite:4paramfamily}.  Towards understanding the spaces of complex Hadamards, numerous methods of construction have been developed which give explicit families of matrices~\cite{cite:karlsson2,cite:karlsson1,cite:evendim}.  Often one imposes additional assumptions on the structure of the matrices, thereby simplifying the problem and allowing algebraic conditions to be solved explicitly.

To study the dimension of the space of complex Hadamards near a fixed matrix $H$, the defect $d(H) \in \mathbb{N}$ can be computed~\cite{cite:defectdef}.  This quantity (defined in Section \ref{sec:theory}) bounds the dimension of affine orbits stemming from a dephased Hadamard~\cite{cite:defectuse}.  Notably, if $d(H) = 0$, then there exists a neighborhood of $H$ which does not contain any other dephased Hadamards, and $H$ is said to be \textbf{isolated}.  The main contributions of this paper include 1) introducing an equivalent definition of the defect using classical dynamical systems theory, 2) demonstrating through examples how this new framework can be used to investigate the local structure of complex Hadamards and, in particular, 3) constructing several new affine families stemming from known Hadamards.

This paper is organized as follows: in Section~\ref{sec:theory} we motivate and define our methodology.  Section~\ref{sec:4x4} serves to justify and explicate the technique by applying it to the space of $4 \times 4$ Hadamard matrices.  In Section~\ref{sec:examples} we use our approach on a host of Hadamard matrices where the local structure is not fully understood to demonstrate both its versatility and limitations. We end with some discussion of the scope of this new perspective.

\section{Hadamards As Fixed Points of a Flow}
\label{sec:theory}
Fix an integer $d \geq 2$ and consider a dephased matrix of the form 
$$
\setlength{\abovedisplayskip}{10pt}
\setlength{\belowdisplayskip}{10pt}
H_d({\bm \theta}) \doteq \begin{bmatrix} 1 & 1 & 1 & \cdots & 1 \\ 1 & e^{\ii\theta_1} & e^{\ii\theta_2} & \cdots & e^{\ii\theta_{d-1}} \\ 1 & e^{\ii\theta_{d}} & e^{\ii\theta_{d+1}} & \cdots & e^{\ii\theta_{2(d-1)}} \\ \vdots & \vdots & \vdots & \ddots & \vdots \\ 1 & e^{\ii\theta_{(d-2)(d-1)+1}} & e^{\ii\theta_{(d-2)(d-1)+2}} & \cdots & e^{\ii\theta_{(d-1)^2}} \end{bmatrix},
$$
depending on ${\bm \theta} \doteq \left[\theta_1, \theta_2, \ldots, \theta_{(d-1)^2}\right]$ with $\theta_i \in [0,2\pi)$.  If $H_d({\bm \theta})$ is a dephased Hadamard, then $H_d({\bm \theta}) H_d({\bm \theta})^* = d I_d$.  Naturally this imposes conditions on the allowed phases in the entries of $H_d({\bm \theta})$. In particular, ${\bm \theta}$ must be chosen to satisfy $d(d-1)$ equations stemming from the requirement that the off-diagonal entries of $H_d({\bm \theta}) H_d({\bm \theta})^*$ must be identically 0:
\begin{equation}
	[H_d({\bm \theta}) H_d({\bm \theta})^*]_{i,j} = \begin{cases} 0, & \mbox{if } i \neq j \\ d, & \mbox{if } i=j \end{cases}.
	\label{eq:potentialdef}
\end{equation}
Using the equations in (\ref{eq:potentialdef}) we define a scalar potential $\mathcal{V}_d: \mathbb{R}^{(d-1)^2} \rightarrow \mathbb{R}$, which can be thought of as measuring the extent of the failure of a matrix to be Hadamard, by
\begin{equation}
\mathcal{V}_d({\bm \theta}) \doteq \sum_{i\neq j}^{d}\abs{\left[H_d({\bm \theta}) H_d({\bm \theta})^*\right]_{i,j}}^2.
	\label{eq:gradientdef}
\end{equation}
\noindent Observe, $\mathcal{V}_d({\bm \theta})$ vanishes exactly when $H_d({\bm \theta})$ is a complex Hadamard matrix.  By computing the negative gradient of $\mathcal{V}_d$, we define a gradient system of ordinary differential equations,
\begin{equation}
\Phi_d({\bm \theta}) \doteq -\nabla\mathcal{V}_d,
\label{eq:gradsys}
\end{equation}

\noindent whose stationary points are the dephased complex Hadamard matrices.  Equation~(\ref{eq:gradsys}) can be thought of as defining a flow on the $(d-1)^2$-torus, $\textbf{T}^{(d-1)^2}$, since only the \textbf{core} -- the $(d-1) \times (d-1)$ lower-right submatrix -- is allowed to vary (i.e. we insist that the matrices $H_d({\bm \theta})$ be dephased).  Note that this system does not take into account permutation equivalences, and so the set to which $\Phi_d({\bm \theta})$ converges is not the space of inequivalent Hadamards, but rather the space of inequivalent Hadamards together with all copies of this space derived from permutations of the cores of its members.

Let $\textbf{0}$ be a fixed point of the nonlinear system 
\begin{equation}
\frac{d\textbf{x}}{dt} = A \textbf{x} + F(\textbf{x}), \textbf{x} \in \mathbb{R}^N,
\label{eq:nonlingensys}
\end{equation}
\noindent where $A$ is an $N \times N$ matrix and $F$ is $C^r$ in a neighborhood of $\textbf{0}$.  Then the eigenvalues of $A$ can be used to determine the local dynamics of the system near the origin.  In particular, there exist stable ($E^s$), unstable ($E^u$), and center ($E^c$) subspaces spanned by the generalized eigenvectors corresponding to the eigenvalues of $A$ with negative, positive, and zero real-parts respectively.  The generalized eigenvectors spanning the stable and unstable subspaces are tangent at the origin to the stable and unstable manifolds ($W^s$ and $W^u$): invariant sets with consistent asymptotic behavior with respect to the origin.  Similarly, the center subspace is tangent to every center manifold, $W^c$: invariant sets on which $F(\textbf{x})$ (the non-linear part of the vector field) governs the dynamics as given by the Center Manifold Theorem \cite{cite:meiss,cite:verhulst}.

\begin{theorem*}
Assume that $A$ has $c$ eigenvalues with real part equal to zero and $N-c$ eigenvalues with negative real part.  Then the system defined by Equation~(\ref{eq:nonlingensys}) can be written in diagonal form
\begin{align*}
\frac{d \textbf{x}}{dt}  & = C \textbf{x} + G(\textbf{x},\textbf{y})\\
\frac{d \textbf{y}}{dt}  & = S \textbf{y} + H(\textbf{x},\textbf{y}),
\end{align*}
\noindent where $\textbf{x} \in \mathbb{R}^c, \textbf{y} \in \mathbb{R}^{N-c}$, $C$ is a square matrix whose eigenvalues all have 0 real part, $S$ is a square matrix whose eigenvalues have negative real part, and $G(\textbf{0}) = H(\textbf{0}) = DG(\textbf{0}) = DH(\textbf{0}) = \textbf{0}$.  Furthermore, for some $\delta > 0$ there exists $h \in C^r(B_{\delta}(\textbf{0}))$ that defines the local center manifold $W^c(\textbf{0}) = \{\left[\textbf{x},\textbf{y}\right] \in \mathbb{R}^N| \textbf{y} = h(\textbf{x}) \text{ for } |\textbf{x}| < \delta\}$ and satisfies
$$
\setlength{\abovedisplayskip}{10pt}
\setlength{\belowdisplayskip}{10pt}
D h(\textbf{x})[C\textbf{x} + G(\textbf{x},h(\textbf{x}))] = S h(\textbf{x})+H(\textbf{x},h(\textbf{x}))
$$
\noindent for $|\textbf{x}| < \delta$.  The flow on the center manifold is governed by the system
$$\setlength{\abovedisplayskip}{10pt}
\setlength{\belowdisplayskip}{10pt}
\frac{d \textbf{x}}{dt} = C \textbf{x} + G(\textbf{x},h(\textbf{x}))
$$ 
\noindent for all $\textbf{x} \in \mathbb{R}^c$ with $\abs{\textbf{x}}<\delta$.
\label{thm:cmthm}
\end{theorem*}

\noindent  Note that the linearized system given by the Center Manifold Theorem has no eigenvalues with positive real part, and thus no unstable subspace or manifold at $\textbf{0}$.  This assumption is not required but is appropriate for us since Hadamard matrices are global minima of the potential (\ref{eq:gradientdef}).

The Center Manifold Theorem provides a means to bound the local dimension of the space of degenerate fixed points, $\mathcal{F}$, containing $\textbf{0}$ since the center manifold at $\textbf{0}$ must contain $\mathcal{F}$.  Thus, by applying the theorem to the gradient system~(\ref{eq:gradsys}), we can estimate the local dimension of the space of dephased complex Hadamards at $H$ by computing the dimension of the center manifold at $H$. 

This construction is reminiscent of the definition of the \textbf{defect} of a $d \times d$ Hadamard matrix $H$ as the dimension of the solution space of the real linear system   
\begin{align*}
	R_{i,1} & = 0,  \text{ for }  1 \leq i \leq d \\
	R_{1,j} & = 0,  \text{ for }  2	 \leq j \leq d \numberthis \label{eq:lindef}\\
	\sum_{k=1}^d [H]_{i,k}[H^*]_{j,k}\left([R]_{i,k}-[R]_{j,k}\right) & = 0, \text{ for }  1 \leq i < j \leq d,
\end{align*}
\noindent where $R \in M^{d \times d}(\mathbb{R})$ is a matrix of variables. The linear system (\ref{eq:lindef}) is derived by considering a matrix $H \circ \text{EXP}(\ii R)$, $\left([\text{EXP}(\ii R)]_{i,j} = e^{\ii [R]_{i,j}}\right)$, and computing the Jacobian of the non-linear system
\begin{align*}	
	R_{i,1} & = 0,  \text{ for }  1 \leq i \leq d \\
	R_{1,j} & = 0,  \text{ for }  2	 \leq j \leq d \numberthis \label{eq:nonlindef} \\
	\sum_{k=1}^d [H]_{i,k}[H^*]_{j,k} e^{\ii \left([R]_{i,k}-[R]_{j,k}\right)} & = 0, \text{ for }  1 \leq i < j \leq d.
\end{align*}
The $d^2+d-1$ equations in (\ref{eq:nonlindef}) follow from the unitary condition, together with the dephased property which must be satisfied if $H \circ \text{EXP}(\ii R)$ is a dephased Hadamard. 

The dimension of the center manifold of system (\ref{eq:gradsys}) at a Hadamard $H$ is actually equal to the defect $d(H)$ since we are linearizing the dephased Hadamard conditions in both cases.  In particular, recall that if the defect of a Hadamard matrix is 0, then it must be isolated.  Likewise, if there does not exist a center subspace of $\Phi_d$ at $H_d({\bm \theta})$, then the stable subspace is $(d-1)^2$-dimensional and all points sufficiently close to $H_d({\bm \theta})$ must flow to it. 

Since a center manifold is not necessarily (and not usually) comprised entirely of fixed points, the Center Manifold Theorem could improve upon an overestimate of the local dimension of the space of complex Hadamards suggested by the the dimension of the center subspace and the defect.  If $\Phi_d({\bm \theta}) \neq 0$, then $H_d({\bm \theta})$ is not a complex Hadamard matrix and therefore any flow on the center manifold, as slow as it may be, will shrink the bound given by the defect.

A center manifold reduction is typically accomplished by performing a change of coordinates on the system into eigen-coordinates so that the center manifold can be written as a graph over the center subspace.  This becomes impractical for a high-dimensional systems since the process first requires diagonalization of (large) symbolic matrices.  We have adopted an alternative method in which the center manifold, $W^c$, is written as an embedding over the $c$-dimensional center subspace, $E^c = $ span($\textbf{v}_1,\ldots,\textbf{v}_c$)~\cite{cite:coordfreecmr}.  One begins by writing  
$$
\setlength{\abovedisplayskip}{10pt}
\setlength{\belowdisplayskip}{10pt}
W^c = \textbf{X}(t_1,\ldots,t_c) = t_1 \textbf{v}_1 + \ldots + t_c \textbf{v}_c + \textbf{w}(t_1,\ldots,t_c),
$$
and expands $\textbf{w}(t_1,\ldots,t_c) \in (E^c)^{\perp}$ in a Taylor expansion by repeated differentiation of the vector field.  By observing that the center manifold is an invariant set (i.e. the flow at a point on the $W^c$ is tangent to $W^c$ there) one derives  
\begin{equation}
	f(\textbf{X}(t_1,\ldots,t_c)) = \alpha_1(t_1,\ldots,t_c) \frac{\partial \textbf{X}}{\partial t_1} + \ldots +\alpha_c(t_1,\ldots,t_c) \frac{\partial \textbf{X}}{\partial t_c},
\label{eq:tangency}
\end{equation}
for some real-valued functions $\alpha_i:\mathbb{R}^c \rightarrow \mathbb{R}$, which can be shown to be the time-rates-of-change of the embedding parameters, $t_1, \ldots, t_c$.  More precisely, $\alpha_i = dt_i/dt$ for $i=1,\ldots, c$.  As one expands $\textbf{w}(t_1,\ldots,t_c)$, they also approximate each $\alpha_i(t_1,\ldots,t_c)$ and thus the flow on the center manifold.  Notice, if flow is nowhere present on the center manifold, then $\textbf{X}(t_1,\ldots,t_c)$ represents a local embedding of dephased Hadamards and so this framework yields a series approximation of the manifold of interest.

The principal computational limitation is the memory requirements of storing high-order tensors, which are needed if one wishes to expand the center manifold to high orders and which grow exponentially with order.  This is exaggerated for large matrices, since the base of the exponential growth depends on the order of the Hadamard.  Although identification of flow on the center manifold may not require expansion to high orders -- as shown by example in Section~\ref{sec:4x4} -- it should be noted that the \textit{absence} of flow at any finite order cannot guarantee that flow is not present at some higher order. 

\section{Explanatory Example}
\label{sec:4x4}
It is known that every $4 \times 4$ Hadamard matrix is equivalent to a member of the continuous one-parameter family of inequivalent Hadamards~\cite{cite:4x4hadams},
$$
\setlength{\abovedisplayskip}{10pt}
\setlength{\belowdisplayskip}{10pt}
F_4^{(1)}(a) \doteq \begin{bmatrix}1 & 1 & 1 & 1 \\ 1 & \ii e^{\ii a} & -1 & -\ii e^{\ii a} \\ 1 & -1 & 1 & -1 \\ 1 & -\ii e^{\ii a} & -1 & \ii e^{\ii a}\end{bmatrix}.
$$
\noindent Moreover, for all but one choice of parameter $a \in [0,\pi]$, the defect of $F_4^{(1)}(a)$ coincides with the dimension of this topological circle of inequivalent Hadamards.  However, for $a = \pi/2$, when the Hadamard is real, the defect jumps from 1 to 3.  Thus, this is an example where we know that the defect overestimates the local freedom of inequivalent Hadamards.  This section serves as a illustrative example of the methodology developed in Section~\ref{sec:theory}. In particular we use the center manifold reduction to show what is already known: The space of inequivalent Hadamards near $F_4^{(1)}(\pi/2)$ is one-dimensional, despite what the defect there might suggest.

For the remainder of this section we will denote the one-parameter family of inequivalent $4 \times 4$ Hadamards as $F(a) \doteq F_4^{(1)}(a)$.  We will interchangeably refer to elements of this space as either the matrix $F(a)$, for a particular $a \in [0,\pi/2]$, or the vector 
$$
\setlength{\abovedisplayskip}{10pt}
\setlength{\belowdisplayskip}{10pt}
 {\bm \theta}(a) \doteq \left[a-\frac{\pi}{2},\pi,a+\frac{\pi}{2},\pi,0,\pi,a+\frac{\pi}{2},\pi,a-\frac{\pi}{2}\right].
$$ 
  
Derivation of the gradient system $\Phi_4({\bm \theta})$ (Equation~\ref{eq:gradsys}) begins with the matrices
$$
\setlength{\abovedisplayskip}{10pt}
\setlength{\belowdisplayskip}{10pt}
\renewcommand{\arraystretch}{0.7}
	H_4({\bm \theta}) \doteq \begin{bmatrix}1 & 1 & 1 & 1 \\ 1 & e^{\ii \theta_1} & e^{\ii \theta_2} & e^{\ii \theta_3} \\ 1 & e^{\ii \theta_4} & e^{\ii \theta_5} & e^{\ii \theta_6} \\ 1 & e^{\ii \theta_7} & e^{\ii \theta_8} & e^{\ii \theta_9} \end{bmatrix}.
$$ 

\begin{figure}[htbp]
	\centering
		\includegraphics[width=.75\textwidth]{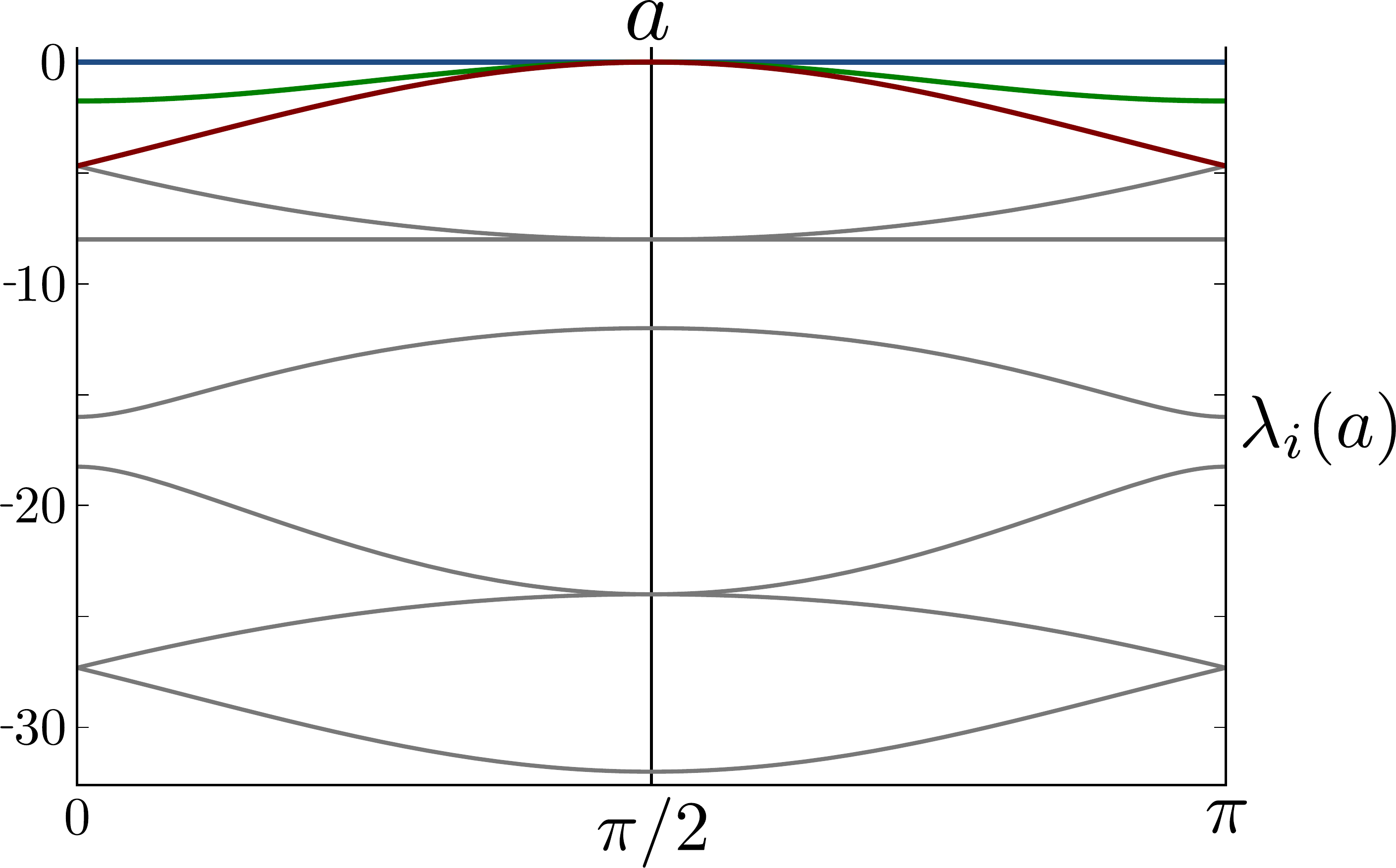}
		\caption{Plot of the eigenvalues of the linearization of $\Phi_4$ at ${\bm \theta}(a)$ for $a \in [0,\pi]$.  $\lambda_1(a)$ (blue), $\lambda_3(a)$ (green), and $\lambda_6(a)$ (red) simultaneously vanish at $a = \pi/2$, while all other eigenvalues (gray) are strictly negative for all $a \in [0,\pi]$.}
	\label{fig:eigenvaluesdim4}
\end{figure}

\noindent With the aid of a computer algebra system, we compute explicit formulae for the eigenvalues of the Jacobian matrix, $D\Phi_4|_{{\bm \theta}(a)}$, as functions of the parameter $a \in [0,\pi]$:  
\setlength{\arraycolsep}{3pt}
$$
\setlength{\abovedisplayskip}{10pt}
\setlength{\belowdisplayskip}{10pt}
	D\Phi_4|_{{\bm \theta}(a)} = \renewcommand{\arraystretch}{.8}
	 \small 4 \begin{bmatrix} -3 & 1& 1& 1& -\sin a& -1& 1& -1& 1\\ 
                 1 & -3& 1& -\sin a& 1& \sin a& -1& 1& -1\\ 
			     1 & 1& -3& -1& \sin a& 1& 1& -1& 1 \\ 
                 1& -\sin a& -1& -3& 1& 1& 1& \sin a& -1\\ 
				-\sin a& 1&  \sin a& 1& -3& 1& \sin a& 1& -\sin a\\ 
                -1& \sin a& 1& 1& 1& -3& -1& -\sin a& 1\\ 
                 1& -1& 1& 1& \sin a& -1& -3& 1& 1\\ 
				-1& 1& -1& \sin a& 1& -\sin a& 1& -3& 1\\ 
				1& -1&  1& -1& -\sin a& 1& 1& 1& -3
															  \end{bmatrix}.
$$
\noindent  The characteristic polynomial of $D\Phi_4|_{{\bm \theta}(a)}$ is 
\begin{align*}
	p(\lambda;a) = -\lambda [\lambda + 8] & [\lambda^2  + (32 - 8\sin a) \lambda + 128(1-\sin a)]\\
	            & [\lambda^2  + (32 + 8\sin a) \lambda + 128(1+\sin a)] \\
							& [\lambda^3  + 36\lambda^2 + (352- 64\sin^2 a)\lambda  + 512(1 - \sin^2 a)].
\end{align*}
\noindent Let	$\lambda_1(a) = 0$, $\lambda_2(a)  = -8$, and $\lambda_3(a)$, $\lambda_4(a)$ and $\lambda_5(a)$ equal the three real roots of the cubic factor
$$
\setlength{\abovedisplayskip}{10pt}
\setlength{\belowdisplayskip}{10pt}
c(\lambda;a) \doteq \lambda^3  + 36\lambda^2 + (352- 64\sin^2 a)\lambda  + 512(1 - \sin^2 a),
$$
\noindent $\lambda_6(a)$ and $\lambda_7(a)$ the roots of the quadratic 
$$
\setlength{\abovedisplayskip}{10pt}
\setlength{\belowdisplayskip}{10pt}
q_1(\lambda;a) \doteq \lambda^2  + (32 - 8\sin a) \lambda + 128(1-\sin a),
$$ 
\noindent and $\lambda_8(a)$ and $\lambda_9(a)$ the roots of the quadratic
$$\\
\setlength{\abovedisplayskip}{10pt}
\setlength{\belowdisplayskip}{10pt}
q_2(\lambda;a) \doteq \lambda^2  + (32 + 8\sin a) \lambda + 128(1+\sin a).
$$
\noindent The constant terms of $c(\lambda; a)$ and $q_1(\lambda;a)$ vanish precisely at $a=\pi/2$, increasing the multiplicity of the root $\lambda = 0$, of $p(\lambda;a)$, to three there.  A plot of the eigenvalues of $D\Phi_4|_{{\bm \theta}(a)}$ is given in Figure~\ref{fig:eigenvaluesdim4}.

As stated, for every parameter value (other than $a = \pi/2$) there are exactly 8 negative eigenvalues and 1 eigenvalue equal to 0. The latter corresponds to the one-dimensional manifold of fixed points parametrized by $a$, as its eigenvector is the tangent vector to the manifold $F_4^{(1)}$ embedded in $\textbf{T}^9$,
$$
\setlength{\abovedisplayskip}{10pt}
\setlength{\belowdisplayskip}{10pt}
\textbf{v}_1 \doteq \left[1,0,1,0,0,0,1,0,1\right].
$$
At $a=\pi/2$, $\lambda_3$ and $\lambda_6$ also vanish, giving rise to a three-dimensional center manifold spanned by $\textbf{v}_1$ and the eigenvectors
$$
\setlength{\abovedisplayskip}{10pt}
\setlength{\belowdisplayskip}{10pt}
\textbf{v}_2 \doteq \left[0,0,0,1,1,0,1,1,0\right] \text{ and } \textbf{v}_3 \doteq \left[0,1,1,0,1,1,0,0,0\right].
$$  
We anticipate nonlinear flow at all nearby points off of the lines spanned by each $\textbf{v}_1$, $\textbf{v}_2$ and $\textbf{v}_3$ since the space of inequivalent Hadamards is one-dimensional at $F(\pi/2)$.  Upon expanding the center manifold as an embedding over $E^c = \text{span}(\textbf{v}_1,\textbf{v}_2,\textbf{v}_3)$,
$$
\setlength{\abovedisplayskip}{10pt}
\setlength{\belowdisplayskip}{10pt}
	\textbf{X}(t_1,t_2,t_3) \doteq t_1 \textbf{v}_1 + t_2 \textbf{v}_2 + t_3 \textbf{v}_3 + \textbf{w}(t_1,t_2,t_3),
$$
\noindent we derive cubic approximations of the time-rates-of-change of the embedding parameters:
\begin{align*}
	\dot{t_1} & = -\frac{20}{9}t_1 {t_2}^2 - \frac{20}{9}t_1 {t_3}^2 + \frac{4}{9} t_2{t_1}^2 + \frac{4}{9}t_2 {t_3}^2 + \frac{4}{9}t_3 {t_1}^2 + \frac{4}{9}t_3 {t_2}^2  \\
	\dot{t_2} & = -\frac{20}{9}t_2 {t_1}^2 - \frac{20}{9}t_2 {t_3}^2 + \frac{4}{9} t_1{t_2}^2 + \frac{4}{9}t_1 {t_3}^2 + \frac{4}{9}t_3 {t_1}^2 + \frac{4}{9}t_3 {t_2}^2  \numberthis \label{eqn:timerates} \\
	\dot{t_3} & = -\frac{20}{9}t_3 {t_1}^2 - \frac{20}{9}t_3 {t_2}^2 + \frac{4}{9} t_1{t_2}^2 + \frac{4}{9}t_1 {t_3}^2 + \frac{4}{9}t_2 {t_1}^2 + \frac{4}{9}t_2 {t_3}^2.
\end{align*}
Thus, the motion of the point $\textbf{X}(t_1,t_2,t_3)$ on $W^c$ is governed by (\ref{eqn:timerates}).  

Because every nonzero cubic-term in (\ref{eqn:timerates}) is a mixed monomial, setting any two embedding parameters to zero will result in no flow.  For example, setting $t_2 = t_3 = 0$ amounts to choosing a point $\textbf{X}(t_1,0,0) \in F_4^{(1)}$ on the manifold of fixed points (since moving in the direction of $\textbf{v}_1$ amounts to varying the parameter $a$). Any solution trajectory of (\ref{eqn:timerates}) converges to a point on one of the axes, where two of the embedding parameters vanish.

The important point is that there {\em is} flow for any choice of embedded point $\textbf{X}(t_1,t_2,t_3)$ not directly over the lines spanned by $\textbf{v}_1$, $\textbf{v}_2$ or $\textbf{v} _3$ (i.e. on an axis in $t_1-t_2-t_3$ space).  Therefore, the local dimension of the space of inequivalent Hadamards at $F(\pi/2)$ cannot be greater than one.

Recall that $\Phi_4$ does not converge to the space of inequivalent Hadamards, but rather to the superset containing all row and column permutations of the core of dephased $4 \times 4$ Hadamards.  Let $P_r(i,j)$ and $P_c(i,j)$ be the $4 \times 4$ permutation matrices which act to swap rows $i$ and $j$ and columns $i$ and $j$ respectively.  There are exactly five unique row and column permutations of $F(a)$ which, for some choice of parameter $a$, are again equal to the matrix $F(\pi/2)$.  In particular,
\begin{align*}
 F(\pi/2) & = F(3\pi/2) P_c(2,4) \\
					& = P_r(2,4) F(3\pi/2) \\
					& = P_r(2,4) F(\pi/2) P_c(2,4) \\
          & = P_r(2,3) F(\pi/2) P_c(3,4) \\
					& = P_r(3,4) F(\pi/2) P_c(2,3)
\end{align*}
Any such core-permutation amounts to a permutation of the the coordinates ${\bm \theta} = [\theta_1, \ldots, \theta_9]$ of $\textbf{T}^9$.  For example, 
$$
\setlength{\abovedisplayskip}{10pt}
\setlength{\belowdisplayskip}{10pt}
P_r(2,3) H_4({\bm \theta}) P_c(3,4) = \renewcommand{\arraystretch}{0.7}
\begin{bmatrix}1 & 1 & 1 & 1 \\ 
						                                           1 & e^{\ii \theta_4} & e^{\ii \theta_6} & e^{\ii \theta_5} \\ 
                                                       1 & e^{\ii \theta_1} & e^{\ii \theta_3} & e^{\ii \theta_2} \\ 
																											 1 & e^{\ii \theta_7} & e^{\ii \theta_9} & e^{\ii \theta_8} \end{bmatrix},
$$
\noindent corresponds to the permutation 
$$
\setlength{\abovedisplayskip}{10pt}
\setlength{\belowdisplayskip}{10pt}
\begin{aligned}
\sigma_2 \doteq (14)(26)(35)(7)(89) \\ 
[\theta_1,\theta_2,\theta_3,\theta_4, \theta_5,\theta_6,\theta_7,\theta_8,\theta_9] \mapsto [\theta_4,\theta_6,\theta_5,\theta_1,\theta_3,\theta_2,\theta_7,\theta_9,\theta_8].
\end{aligned}
$$  
\noindent Let $\sigma_3 \doteq (12)(3)(48)(57)(69)$; the permutation of the coordinates resulting from the action of $P_r(3,4)$ and $P_c(2,3)$.  These permutations act on the vector $\textbf{v}_1$, which we recall is tangent to $F(a)$ at $a =\pi/2$, in the following ways:
\begin{align*}
\setlength{\abovedisplayskip}{10pt}
\setlength{\belowdisplayskip}{10pt}
	\textbf{v}_1 = \left[1,0,1,0,0,0,1,0,1\right] & \xmapsto{\sigma_2} \left[0,0,0,1,1,0,1,1,0\right] = \textbf{v}_2, \text{ and } \\
	\textbf{v}_1 = \left[1,0,1,0,0,0,1,0,1\right] & \xmapsto{\sigma_3} \left[0,1,1,0,1,1,0,0,0\right] = \textbf{v}_3,
\end{align*}

\noindent while $v_1$ remains fixed under $P_r(2,4)$ and $P_c(2,4)$. We see that there are five total (three distinct) directions of fixed-point degeneracy at $F(\pi/2)$ caused by the permutations of its core.  We conclude that the three-dimensional center manifold emerges at the real Hadamard to account for three copies of the space of dephased, permutation-equivalent Hadamards intersecting here. 

\begin{figure}[htbp]
	\centering
		\includegraphics[width=1.0\textwidth]{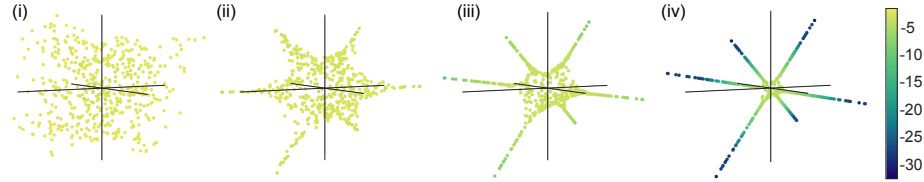}
		\caption{Snapshots of 500 initial phases -- drawn from $\mathbb{R}^{9}$ uniformly at random from a neighborhood of the core phases corresponding to $F(\pi/2)$ -- as they evolve under the flow defined by $\Phi_4({\bm \theta})$, at times (i) 5, (ii) 20, (iii) 70, and (iv) 500.  Each point cloud has been projected onto its top three principal components, and each point ${\bm \theta}$ is colored by $\log_{10}$ of the magnitude of the vector field $\Phi_4({\bm \theta})$.}
	\label{fig:flowf41}
\end{figure}

The local geometric consequence of this result can be seen quite convincingly by directly visualizing the flow of core phases taken from a neighborhood of $F(\pi/2)$.  Figure~\ref{fig:flowf41} shows snapshots of the evolution of a point cloud of 500 initial conditions drawn uniformly at random from a neighborhood of $F(\pi/2)$. The point cloud appears to be converging to three lines intersecting at $F(\pi/2)$, as expected in light of (\ref{eqn:timerates}).  An ancillary animation reflecting the flow of $\Phi_4({\bm \theta})$ in $\mathbb{R}^9$ near $F(\pi/2)$ is provided.

\section{Applications}
\label{sec:examples}
\label{sec:6x6}
The smallest order for which classification of complex Hadamards is incomplete is order six.  Numerical searches, analysis of known families, and a general method of construction due to Sz{\"o}ll{\H{o}}si, which depends on four free parameters, all support the conjecture that Sz{\"o}ll{\H{o}}si's generic four-parameter family and the isolated matrix $S_6^{(0)}$ capture all inequivalent $6 \times 6$ Hadamards~\cite{cite:4paramfamily}.  This is evidenced by calculation of the defect at both exact and numerical Hadamards generated from known families and by random searches: save for $S_6^{(0)}$ the defect is always found to be four.  We further support the conjecture by expanding the flow on the center manifold of $\Phi_6$ at selected Hadamards without ever encountering non-zero coefficients in the Taylor expansion of the time-rates-of-change of the embedding parameters.  

\noindent \textbf{The Affine Fourier Family} Stemming from the Fourier matrix
$$
\setlength{\abovedisplayskip}{10pt}
\setlength{\belowdisplayskip}{10pt}
\renewcommand{\arraystretch}{0.7}
	F_6 = \begin{bmatrix} 
			1 & 1 & 1 & 1 & 1 & 1 \\ 
			1 & w & w^2 & w^3 & w^4 & w^5 \\
			1 & w^2 & w^4 & 1 & w^2 & w^4 \\
			1 & w^3 & 1 & w^3 & 1 & w^3 \\
			1 & w^4 & w^2 & 1 &  w^4 & w^2 \\
			1 & w^5 & w^4 & w^3 & w^2 & w 
		  \end{bmatrix},
$$
(where $w = e^{2\pi \ii/6}$) are two, two-parameter affine orbits of Hadamards
\begin{align*}
\setlength{\abovedisplayskip}{10pt}
\setlength{\belowdisplayskip}{10pt}
	F_6^{(2)}(a,b) & := F_6 \circ \text{EXP}\left(\ii R(a,b)\right) \\ 
	F_6^{(2)}(a,b)^{\text{T}} & :=  F_6 \circ \text{EXP}\left(\ii R(a,b)^{\text{T}}\right),
\end{align*}
where 

$$
\setlength{\abovedisplayskip}{10pt}
\setlength{\belowdisplayskip}{10pt}
\renewcommand{\arraystretch}{0.5}
	R(a,b) = \begin{bmatrix} 
	\bullet & \bullet & \bullet & \bullet & \bullet & \bullet \\
	\bullet & a & b & \bullet & a & b \\
	\bullet & \bullet & \bullet & \bullet & \bullet & \bullet \\
	\bullet & a & b & \bullet & a & b \\
	\bullet & \bullet & \bullet & \bullet & \bullet & \bullet \\
	\bullet & a & b & \bullet & a & b \\
	 \end{bmatrix}.
$$ 
Note that $H\circ K$ denotes the Hadamard (entrywise) product of the matrices $H$ and $K$, and the numeric value 0 has been replaced with $\bullet$ to improve readability. 

The defect of $F_6$ is four, and numerical evidence supports the conjecture that there is a four-dimensional non-affine family stemming from it~\cite{cite:expansionmethod}. In kind the center subspace of $\Phi_6$ at $F_6$ (i.e. the kernel of the symmetric matrix $D\Phi_6 \vert_{F_6}$) is four dimensional and is spanned by the vectors
\begin{align*}
	\textbf{v}_1 & = [1, 0, 0, 1, 0, 0, 0, 0, 0, 0, 1, 0, 0, 1, 0, 0, 0, 0, 0, 0, 1, 0, 0, 1, 0]\\
	\textbf{v}_2 & = [0, 1, 0, 0, 1, 0, 0, 0, 0, 0, 0, 1, 0, 0, 1, 0, 0, 0, 0, 0, 0, 1, 0, 0, 1] \\
	\textbf{v}_3 & = [1, 0, 1, 0, 1, 0, 0, 0, 0, 0, 0, 0, 0, 0, 0, 1, 0, 1, 0, 1, 0, 0, 0, 0, 0]\\
	\textbf{v}_4 & = [0, 0, 0, 0, 0, 1, 0, 1, 0, 1, 0, 0, 0, 0, 0, 0, 0, 0, 0, 0, 1, 0, 1, 0, 1] .
\end{align*}
Recalling the coordinates of $\textbf{T}^{25}$ in which $\Phi_6$ is expressed, it is clear that  $\textbf{v}_1$ and $\textbf{v}_2$ span $F_6^{(2)}(a,b)$ and $\textbf{v}_3$ and $\textbf{v}_4$ span $F_6^{(2)}(a,b)^{\text{T}}$. 

We compute the center manifold
$$
\setlength{\abovedisplayskip}{10pt}
\setlength{\belowdisplayskip}{10pt}
\textbf{X}(t_1,t_2,t_3,t_4) = \textbf{w}(t_1,t_2,t_3,t_4) + \sum_{i=1}^{4}t_i \textbf{v}_i,
$$
as an embedding and, as before, expand the functions $\alpha_i(t) = \dot{t_i}$ in power series.  In support of the conjecture that there exists a four-dimensional manifold of complex Hadamards passing through $F_6$, all partial derivatives of each $\alpha_i$ through fifth order are found to vanish.  This means that if the four-dimensional center manifold stemming from $F_6$ does not consist entirely of fixed points, the flow near $F_6$ must be very slow.  This may be interpreted in the following way: points on the local center manifold are very close to being Hadamard, although they may not be.  If the local manifold is four-dimensional, our expansion of $\textbf{X}(t_1,t_2,t_3,t_4)$ gives a series approximation of the space of dephased Hadamards near $F_6$.  Again we cannot rule out the possibility that a higher-order expansion might reveal flow on the center manifold.

\vspace{1mm}

\begin{figure}[htbp]
	\centering
		\includegraphics[width=.75\textwidth]{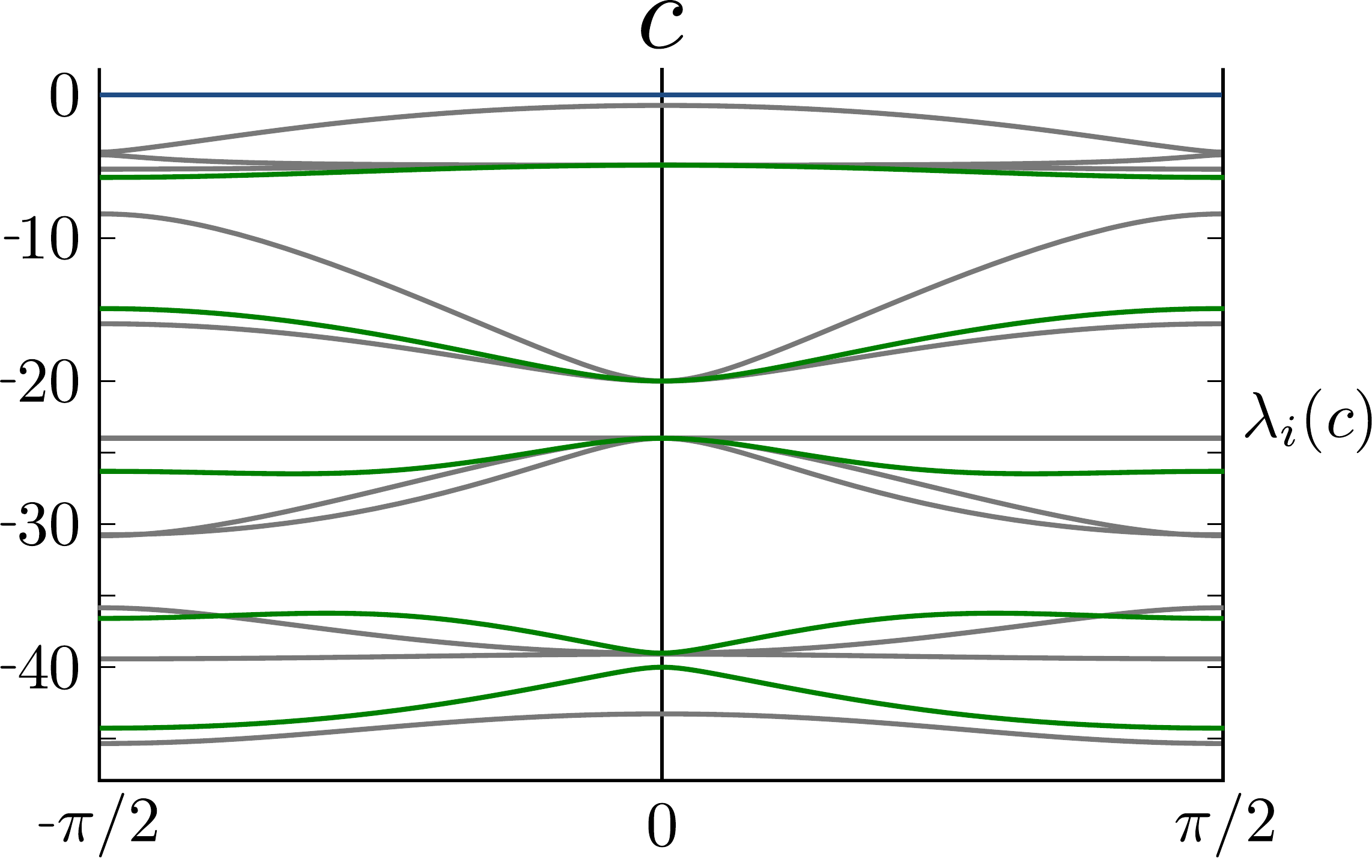}
		\caption{Plot of the 25 eigenvalues of $D\Phi_6\vert_{D_6(c)}$ for $c \in [-\pi/2,\pi/2]$.  The zero eigenvalue (blue) has multiplicity four, the roots of $f_1(\lambda;c)$ (green) have multiplicity two, and all other eigenvalues (gray) are simple.}
	\label{fig:eigenvaluesdim6}
\end{figure}

\noindent \textbf{The affine family} \bm{$D_6^{(1)}$} Another maximal affine order-6 family, this one found by Di{\c{t}}{\u{a}}~\cite{cite:knownfamilies2}, stems from the symmetric matrix 
$$
\setlength{\abovedisplayskip}{10pt}
\setlength{\belowdisplayskip}{10pt}
\renewcommand{\arraystretch}{0.7}
	D_6 = \begin{bmatrix} 
			1 & 1 & 1 & 1 & 1 & 1 \\ 
			1 & -1 & \ii & -\ii & -\ii & \ii \\
			1 & \ii & -1 & \ii & -\ii & -\ii \\
			1 & -\ii & \ii & -1 & \ii & -\ii \\
			1 & -\ii & -\ii & \ii &  -1  & \ii \\
			1 & \ii & -\ii & -\ii  & \ii  &  -1
		  \end{bmatrix}.
$$
One representative of the five, permutation-equivalent families stemming from $D_6$ is $D_6(c) := D_6^{(1)}(c) = D_6 \circ \text{EXP}\left(\ii R(c)\right)$, where
$$
\setlength{\abovedisplayskip}{10pt}
\setlength{\belowdisplayskip}{10pt}
\renewcommand{\arraystretch}{0.5}
	R(c) = \begin{bmatrix} 
	\bullet & \bullet & \bullet & \bullet & \bullet & \bullet \\
	\bullet & \bullet & \bullet & \bullet & \bullet & \bullet \\
	\bullet & \bullet & \bullet & c & c & \bullet \\
	\bullet & \bullet & -c & \bullet & \bullet & -c \\
	\bullet & \bullet & -c & \bullet & \bullet & -c\\
	\bullet & \bullet & \bullet & c & c & \bullet 
	 \end{bmatrix},
$$ 
Computing $D\Phi_6\vert_{D_6(c)}$ gives the linearized flow explicitly in terms of $c$.  The characteristic polynomial of $D\Phi_6\vert_{D_6(c)}$ is found to be
$$
\setlength{\abovedisplayskip}{10pt}
\setlength{\belowdisplayskip}{10pt}
p(\lambda;c) = -\lambda^4(\lambda+24)f_1^2f_2f_3,
$$ 
where
\begin{align*}
	\begin{split} f_1(\lambda;c) = & \; 32\lambda \left(\lambda^2+64\lambda+1024 \right) \cos (2 c)+512 \lambda \cos (4 c)\\ & + \lambda^5+128 \lambda^4+6096 \lambda^3+131840 \lambda^2+1241600 \lambda+3686400
	  \end{split} 
\\
		\begin{split} f_2(\lambda;c) = & \; 256 \lambda \cos (2 c)+\lambda^4+88 \lambda^3+2608 \lambda^2+29312 \lambda+92160
		\end{split} 
\\
		\begin{split} f_3(\lambda;c) = & \; 64 \left(\lambda^4+76 \lambda^3+1408 \lambda^2-2176 \lambda-36864\right) \cos (2 c)\\ & + \lambda^6+132 \lambda^5+6448 \lambda^4+142272 \lambda^3+1386496 \lambda^2+5140480 \lambda+5308416.
		\end{split}
\end{align*}
The factor $\lambda^4$ in $p(\lambda;c)$ guarantees that the center subspace at $D_6(c)$ is at least four dimensional for all values of $c$.  A plot of the eigenvalues of $D\Phi_6\vert_{D_6(c)}$ -- given in Figure~\ref{fig:eigenvaluesdim6} -- suggests that the stable subspace is 21 dimensional for all choices of $c$. In fact, careful consideration of the factors that depend on $c$ reveals that the center subspace is exactly four dimensional for all $c$ since $\lambda=0$ is never a root of $f_1$, $f_2$, or $f_3$. This follows from the observation that $\lambda$ is not a divisor of $f_1$ or $f_2$ for any $c$ since the constant terms in these factors do not depend on $c$.  In $f_3$, the constant (in $\lambda$) term is $-2359296 \cos(2c) + 5308416$ which cannot be made to vanish with $c \in \mathbb{R}$.  This proves that $d \left(D_6^{(1)}(c)\right) = 4$ for every member of this affine family. 

Permutations of the core of $D_6(c)$ give four other families which are permutation equivalent.  These permutations act on the coordinates of the vector tangent to the curve $D_6(c)$ at $D_6(0)$,
$$
\setlength{\abovedisplayskip}{10pt}
\setlength{\belowdisplayskip}{10pt}
\textbf{v}_1  = [0, 0, 0, 0, 0, 0, 0, 1, 1, 0, 0, -1, 0, 0, -1, 0, -1, 0, 0, -1, 0, 0, 1, 1, 0],
$$ 
to give
\begin{align*}
\setlength{\abovedisplayskip}{10pt}
\setlength{\belowdisplayskip}{10pt}
	\textbf{v}_2 & = [0, 0, 0, 1, 1, 0, 0, 0, 0, 0, 0, 0, 0, 1, 1, -1, 0, -1, 0, 0, -1, 0, -1, 0, 0]\\
	\textbf{v}_3 & = [0, -1, 0, -1, 0, 1, 0, 0, 0, 1, 0, 0, 0, 0, 0, 1, 0, 0, 0, 1, 0, -1, 0, -1, 0]\\
	\textbf{v}_4 & = [0, 1, 1, 0, 0, -1, 0, 0, -1, 0, -1, 0, 0, -1, 0, 0, 1, 1, 0, 0, 0, 0, 0, 0, 0]\\
	\textbf{v}_5 & = [0, 0, -1, 0, -1, 0, 0, -1, 0, -1, 1, 1, 0, 0, 0, 0, 0, 0, 0, 0, 1, 1, 0, 0, 0].
\end{align*}
Since $-\textbf{v}_5 = \textbf{v}_1+\textbf{v}_2+\textbf{v}_3+\textbf{v}_4$, we take $\{\textbf{v}_1, \textbf{v}_2, \textbf{v}_3, \textbf{v}_4\}$ as a natural choice of basis for the center subspace at $D_6$.  If flow exists on the center manifold, our choice of basis ensures that it will present itself as nonzero coefficients of the mixed monomials in the Taylor expansions of $\dot{t_1},\ldots,\dot{t_4}.$  Unsurprisingly, we did not encounter any nonzero coefficients in these expansions, reinforcing the conjecture that the space of dephased Hadamards near $D_6$ is four dimensional.

\vspace{1mm}

\noindent \textbf{Beauchamp and Nicoara's} \bm{$B_9^{(0)}$} Often one is able to obtain explicit eigenvectors and eigenvalues for the linearizations of  $\Phi_i({\bm \theta})$ and explicit coefficients in the Taylor expansions of the center manifold and the time-rates-of-change of the embedding parameters.  If numerical computations are undertaken, numerical approximations of the time-rates-of-change of the embedding parameters are determined.  If these approximate coefficients are bounded away from 0, one can deduce that there exists non-linear flow on parts of the center manifold, even without exact knowledge of that flow. 

It is known that the defect of the matrix 
$$
\setlength{\abovedisplayskip}{10pt}
\setlength{\belowdisplayskip}{10pt}
B_9^{(0)} = \renewcommand{\arraystretch}{.6}
            \begin{bmatrix} 1 &       1      &      1     &     1      &     1      &      1      &       1    &      1     &      1      \\
                            1 &      -1      & \epsilon^3 & \epsilon^3 &    -1      & \epsilon^9  & \epsilon^8 & \epsilon^7 & \epsilon    \\ 
														1 &  \epsilon^4  &     -1     & \epsilon^7 & \epsilon   & \epsilon^3  &      -1    & \epsilon^9 & \epsilon^9  \\
														1 &  \epsilon^3  & \epsilon^7 &     -1     & \epsilon   & \epsilon^8  & \epsilon^9 & \epsilon^3 &     -1      \\
														1 &  \epsilon^9  & \epsilon   &     -1     &    -1      & \epsilon^3  & \epsilon^7 & \epsilon^2 & \epsilon^7  \\
														1 &  \epsilon^9  &     -1     &  \epsilon  & \epsilon^3 &      -1     &  \epsilon  & \epsilon^7 & \epsilon^6  \\
														1 &   \epsilon   & \epsilon^7 & \epsilon^9 & \epsilon^6 &  \epsilon   &      -1    &     -1     & \epsilon^3  \\ 	
														1 &  \epsilon^7  & \epsilon^9 & \epsilon^4 & \epsilon^9 &      -1     & \epsilon^3 &     -1     &  \epsilon   \\
														1 &      -1      & \epsilon^2 & \epsilon^9 & \epsilon^7 & \epsilon^7  & \epsilon^3 &  \epsilon  &     -1       \end{bmatrix},
$$
where $\epsilon = e^{2\pi\ii/10}$, is $d\left(B_9^{(0)}\right) = 2$~\cite{cite:isolatedhads}.  Since no family was known that contained $B_{9}^{(0)}$, it was a longstanding open problem to determine if this matrix was actually isolated, despite having a positive defect.  Recently, it was shown that, in fact, the defect arises from a non-affine 2-dimensional family \cite{cite:karlb90} containing it.  Prior to learning of these results, we computed -- to 200 decimal places of accuracy -- numerical approximations of the basis vectors, $\textbf{v}_1$ and $\textbf{v}_2$, which span the center subspace.  We then expanded the center manifold
$\textbf{X}(t_1,t_2) \doteq t_1 \textbf{v}_1 + t_2 \textbf{v}_2 + \textbf{w}(t_1,t_1)$
\noindent as a numerical embedding over the center subspace.  Naturally, the numerical derivatives of the embedding parameters remain zero, as they must at all orders.

\begin{figure}[htbp]
	\centering
		\includegraphics[width=1.0\textwidth]{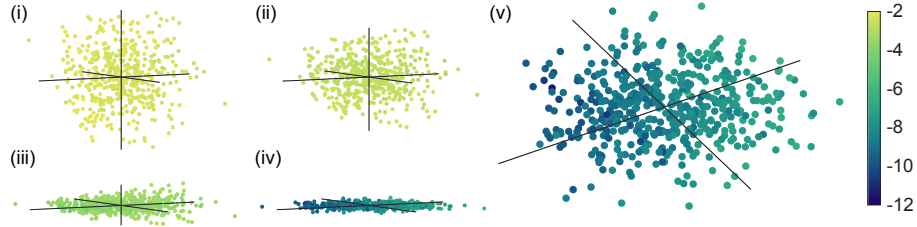}
		\caption{Snapshots of 500 initial phases -- drawn from $\mathbb{R}^{64}$ uniformly at random from a neighborhood of the core phases corresponding to $B_9^{(0)}$ -- as they evolve under the flow defined by $\Phi_9({\bm \theta})$, at times (i) 5, (ii) 20, (iii) 70, and (iv,v) 500.  Each point cloud has been projected onto its top three principal components, and each point ${\bm \theta}$ is colored by $\log_{10}$ of the magnitude of the vector field $\Phi_9({\bm \theta})$.}
	\label{fig:flowd90}
\end{figure}

Visualizing the flow of $\Phi_9({\bm \theta})$ in a neighborhood of $B_9^{(0)}$, which appears to be drawn to a 2-dimensional plane (See Figure~\ref{fig:flowd90} and ancillary files), provides another type of evidence of the conclusions which were proven definitively in \cite{cite:karlb90}. Thus, this serves as an example where the numerical series expansion and numerical integration of the flow was only able to provide evidence for the existence of a center manifold of fixed points.

\vspace{1mm}

\noindent  \textbf{Golay sequence affine family}  $\bm{G_{10}^{(1)}}$ \textbf{\& non-Di{\c{t}}{\u{a}}-type} $\bm{D_{10}^{(3)}}$ In this section we derive two $10 \times 10$ affine families stemming from a member of the one parameter family 
$$
\setlength{\abovedisplayskip}{10pt}
\setlength{\belowdisplayskip}{10pt}
G_{10}^{(1)}(a) = \renewcommand{\arraystretch}{.6} \begin{bmatrix} 
 1 & 1 & 1 & 1 & 1 & 1 & 1 & 1 & 1 & 1 \\
 1 & \ii e^{ \ii a} & \ii e^{ \ii a} & e^{ \ii a} & -\ii e^{ \ii a} & -1 & e^{ \ii a} & -\ii e^{ \ii a} & -e^{ \ii a} & -e^{ \ii a} \\
 1 & -1 & -\ii e^{ \ii a} & -e^{ \ii a} & e^{ \ii a} & \ii & -\ii & e^{ \ii a} & -e^{ \ii a} & \ii e^{ \ii a} \\
 1 & \ii & -\ii & -\ii e^{ \ii a} & \ii e^{ \ii a} & -\ii & \ii & -1 & -e^{ \ii a} & e^{ \ii a} \\
 1 & 1 & \ii & -1 & \ii e^{ \ii a} & \ii & -1 & -\ii & -\ii & -\ii e^{ \ii a} \\
 1 & -\ii e^{ \ii a} & -e^{ \ii a} & e^{ \ii a} & e^{ \ii a} & -1 & -e^{ \ii a} & -e^{ \ii a} & e^{ \ii a} & \ii e^{ \ii a} \\
 1 & \ii & e^{ \ii a} & -e^{ \ii a} & -\ii e^{ \ii a} & -\ii & -1 & \ii e^{ \ii a} & e^{ \ii a} & -e^{ \ii a} \\
 1 & -1 & -\ii & \ii e^{ \ii a} & -e^{ \ii a} & \ii & 1 & -1 & e^{ \ii a} & -\ii e^{ \ii a} \\
 1 & -\ii & \ii & \ii & -e^{ \ii a} & -\ii & -\ii & \ii & -1 & e^{ \ii a} \\
 1 & -\ii & -1 & -\ii & -1 & 1 & \ii & 1 & \ii & -1  \\ 
 \end{bmatrix},
$$
presented here in dephased form, discovered by Lampio \textit{et al}.~\cite{cite:golay}.  The defect of $G_{10}(0) := G_{10}^{(1)}(0)$ and the dimension of the center subspace of $\Phi_{10}$ there is 8, so it may be possible to introduce additional affine parameters.  As we have seen, the center subspace at a Hadamard, $H$, must contain all vectors tangent to dephased families stemming from $H$, affine or otherwise.  For example, the vector $\textbf{v} \in \mathbb{R}^{81}$ with 1's in the coordinates corresponding to the core entries of $G_{10}^{(1)}(a)$ whose phases depend on $a$ (and 0's elsewhere) will be in the center subspace because it is tangent to the line of Hadamards, $G_{10}^{(1)}(a)$. Thus, one can search for affine families by considering linear combinations of vectors in a basis for the center subspace and introducing appropriately-scaled parameters in the phases corresponding to nonzero coordinates.  

Contained in the kernel of $D\Phi_{10}\vert_{G_{10}(0)}$ is the vector 
$$
\setlength{\abovedisplayskip}{10pt}
\setlength{\belowdisplayskip}{10pt}
\renewcommand{\arraystretch}{.6}
\textbf{V} := \begin{bmatrix}
\bullet &\bullet &\bullet &\bullet &\bullet &\bullet &\bullet &\bullet &\bullet &\bullet \\
 \bullet &-1 & \bullet & -1 & -1 & \bullet & \bullet & \bullet & \bullet & -1 \\
 \bullet &\bullet & \bullet & -1 & -1 & \bullet & \bullet & \bullet & \bullet & \bullet \\
 \bullet &\bullet & \bullet & -1 & -1 & \bullet & \bullet & \bullet & \bullet & \bullet \\
 \bullet &\bullet & 1 & \bullet & \bullet & \bullet & \bullet & 1 & \bullet & \bullet \\
 \bullet &\bullet & \bullet & \bullet & \bullet & \bullet & \bullet & \bullet & \bullet & \bullet \\
 \bullet &\bullet & 1 & \bullet & \bullet & 1 & 1 & 1 & \bullet & \bullet \\
 \bullet &\bullet & 1 & \bullet & \bullet & 1 & 1 & 1 & \bullet & \bullet \\
 \bullet &-1 & \bullet & -1 & -1 & \bullet & \bullet & \bullet & \bullet & -1 \\
 \bullet &\bullet & 1 & \bullet & \bullet & \bullet & \bullet & 1 & \bullet & \bullet 
\end{bmatrix},
$$
shown here as the lower-right $9\times 9$ submatrix of $\textbf{V}$ to establish its relationship with the core of $G_{10}(0)$. By computing $G_{10}(0) \circ \text{EXP}\left(a \ii \textbf{V}\right)$ (where $a$ is a free parameter) we uncover a one-parameter affine family; explicitly
$$
\setlength{\abovedisplayskip}{10pt}
\setlength{\belowdisplayskip}{10pt}
M_{10}^{(1)}(a) := \renewcommand{\arraystretch}{.6} \begin{bmatrix}
 1 & 1 & 1 & 1 & 1 & 1 & 1 & 1 & 1 & 1 \\
 1 & \ii e^{-\ii a} & \ii & e^{-\ii a} & -\ii e^{-\ii a} & -1 & 1 & -\ii & -1 & -e^{-\ii a} \\
 1 & -1 & -\ii & -e^{-\ii a} & e^{-\ii a} & \ii & -\ii & 1 & -1 & \ii \\
 1 & \ii & -\ii & -\ii e^{-\ii a} & \ii e^{-\ii a} & -\ii & \ii & -1 & -1 & 1 \\
 1 & 1 & \ii e^{\ii a} & -1 & \ii & \ii & -1 & -\ii e^{\ii a} & -\ii & -\ii \\
 1 & -\ii & -1 & 1 & 1 & -1 & -1 & -1 & 1 & \ii \\
 1 & \ii & e^{\ii a} & -1 & -\ii & -\ii e^{\ii a} & -e^{\ii a} & \ii e^{\ii a} & 1 & -1 \\
 1 & -1 & -\ii e^{\ii a} & \ii & -1 & \ii e^{\ii a} & e^{\ii a} & -e^{\ii a} & 1 & -\ii \\
 1 & -\ii e^{-\ii a} & \ii & \ii e^{-\ii a} & -e^{-\ii a} & -\ii & -\ii & \ii & -1 & e^{-\ii a} \\
 1 & -\ii & -e^{\ii a} & -\ii & -1 & 1 & \ii & e^{\ii a} & \ii & -1 \\
\end{bmatrix}.
$$
Similarly, the null vectors of $D\Phi_{10}\vert_{G_{10}(0)}$, expressed in the cores of
$$
\setlength{\abovedisplayskip}{10pt}
\setlength{\belowdisplayskip}{10pt}
\begin{array}{ccc}
\textbf{U} := \renewcommand{\arraystretch}{.6} \begin{bmatrix}
\bullet & \bullet & \bullet & \bullet & \bullet & \bullet & \bullet & \bullet & \bullet & \bullet \\
 \bullet &\bullet & \bullet & \bullet & \bullet & \bullet & \bullet & \bullet & \bullet & \bullet \\
 \bullet &\bullet & \bullet & 1 & 1 & \bullet & 1 & 1 & 1 & 1 \\
 \bullet &\bullet & \bullet & 1 & \bullet & \bullet & 1 & \bullet & \bullet & \bullet \\
 \bullet &\bullet & \bullet & \bullet & \bullet & \bullet & \bullet & \bullet & \bullet & \bullet \\
 \bullet &\bullet & \bullet & 1 & \bullet & \bullet & 1 & \bullet & \bullet & \bullet \\
 \bullet &\bullet & \bullet & \bullet & \bullet & \bullet & \bullet & \bullet & \bullet & \bullet \\
 \bullet &\bullet & \bullet & 1 & 1 & \bullet & 1 & 1 & 1 & 1 \\
 \bullet &\bullet & \bullet & 1 & \bullet & \bullet & 1 & \bullet & \bullet & \bullet \\
 \bullet &\bullet & \bullet & 1 & \bullet & \bullet & 1 & \bullet & \bullet & \bullet
\end{bmatrix} & \text{ and } &
\textbf{W} := 
\end{array} \renewcommand{\arraystretch}{.6} \begin{bmatrix}
\bullet & \bullet & \bullet & \bullet & \bullet & \bullet & \bullet & \bullet & \bullet & \bullet \\
 \bullet &\bullet & 1 & 1 & \bullet & 1 & 1 & 1 & 1 & \bullet \\
 \bullet &\bullet & 1 & \bullet & \bullet & 1 & \bullet & \bullet & \bullet & \bullet \\
 \bullet &\bullet & \bullet & \bullet & \bullet & \bullet & \bullet & \bullet & \bullet & \bullet \\
 \bullet &\bullet & \bullet & \bullet & \bullet & \bullet & \bullet & \bullet & \bullet & \bullet \\
 \bullet &\bullet & \bullet & \bullet & \bullet & \bullet & \bullet & \bullet & \bullet & \bullet \\
 \bullet &\bullet & 1 & 1 & \bullet & 1 & 1 & 1 & 1 & \bullet \\
 \bullet &\bullet & 1 & \bullet & \bullet & 1 & \bullet & \bullet & \bullet & \bullet \\
 \bullet &\bullet & 1 & \bullet & \bullet & 1 & \bullet & \bullet & \bullet & \bullet \\
 \bullet &\bullet & 1 & \bullet & \bullet & 1 & \bullet & \bullet & \bullet & \bullet
\end{bmatrix},
$$
together reveal the two-parameter family $G_{10}(0) \circ \text{EXP}\left(a \ii \textbf{U}+b \ii \textbf{W}\right)$ 
$$
\setlength{\abovedisplayskip}{10pt}
\setlength{\belowdisplayskip}{10pt}
M_{10}^{(2)}(a,b) := \renewcommand{\arraystretch}{.6} \begin{bmatrix}
 1 & 1 & 1 & 1 & 1 & 1 & 1 & 1 & 1 & 1 \\
 1 & \ii & \ii e^{\ii b} & e^{\ii b} & -\ii & -e^{\ii b} & e^{\ii b} & -\ii e^{\ii b} & -e^{\ii b} & -1 \\
 1 & -1 & -\ii e^{\ii b} & -e^{\ii a} & e^{\ii a} & \ii e^{\ii b} & -\ii e^{\ii a} & e^{\ii a} & -e^{\ii a} & \ii e^{\ii a} \\
 1 & \ii & -\ii & -\ii e^{\ii a} & \ii & -\ii & \ii e^{\ii a} & -1 & -1 & 1 \\
 1 & 1 & \ii & -1 & \ii & \ii & -1 & -\ii & -\ii & -\ii \\
 1 & -\ii & -1 & e^{\ii a} & 1 & -1 & -e^{\ii a} & -1 & 1 & \ii \\
 1 & \ii & e^{\ii b} & -e^{\ii b} & -\ii & -\ii e^{\ii b} & -e^{\ii b} & \ii e^{\ii b} & e^{\ii b} & -1 \\
 1 & -1 & -\ii e^{\ii b} & \ii e^{\ii a} & -e^{\ii a} & \ii e^{\ii b} & e^{\ii a} & -e^{\ii a} & e^{\ii a} & -\ii e^{\ii a} \\
 1 & -\ii & \ii e^{\ii b} & \ii e^{\ii a} & -1 & -\ii e^{\ii b} & -\ii e^{\ii a} & \ii & -1 & 1 \\
 1 & -\ii & -e^{\ii b} & -\ii e^{\ii a} & -1 & e^{\ii b} & \ii e^{\ii a} & 1 & \ii & -1 \\
\end{bmatrix}.
$$

$M_{10}^{(1)}$ and $M_{10}^{(2)}$ were verified to be Hadamard by symbolic computation.  Interestingly, these families and $G_{10}^{(1)}$ are independent in the sense that they do not, together, form a three-parameter affine family.  That being said, a center manifold reduction did not rule out the possibility that they belong to some unifying family that has yet to be discovered.

We applied the same approach to the non-Di{\c{t}}{\u{a}}-type matrix
$$
\setlength{\abovedisplayskip}{10pt}
\setlength{\belowdisplayskip}{10pt}
D_{10} = D_{10}^{(3)}(0,0,0):= \renewcommand{\arraystretch}{.6} \begin{bmatrix} 
 1 & 1 & 1 & 1 & 1 & 1 & 1 & 1 & 1 & 1 \\
 1 & -1 & -\ii & -\ii & -\ii & -\ii & \ii & \ii & \ii & \ii \\
 1 & -\ii & -1 & \ii & \ii & -\ii & -\ii & -\ii & \ii & \ii \\
 1 & -\ii & \ii & -1 & -\ii & \ii & -\ii & \ii & -\ii & \ii \\
 1 & -\ii & \ii & -\ii & -1 & \ii & \ii & -\ii & \ii & -\ii \\
 1 & -\ii & -\ii & \ii & \ii & -1 & \ii & \ii & -\ii & -\ii \\
 1 & \ii & -\ii & -\ii & \ii & \ii & -1 & -\ii & -\ii & \ii \\
 1 & \ii & -\ii & \ii & -\ii & \ii & -\ii & -1 & \ii & -\ii \\
 1 & \ii & \ii & -\ii & \ii & -\ii & -\ii & \ii & -1 & -\ii \\
 1 & \ii & \ii & \ii & -\ii & -\ii & \ii & -\ii & -\ii & -1 \\
\end{bmatrix},
$$
whose membership in a three-parameter affine family $D_{10}^{(3)}$ was already established by Sz{\"o}ll{\H{o}}si~\cite{cite:paramchm}. Notice that the defect of $D_{10}$ is 16, and so there may be more parameters which can be introduced to the family $D_{10}^{(3)}(a,b,c)$.  Towards identifying these additional parameters, we searched for and found a basis, $\{\textbf{V}_1,\ldots,\textbf{V}_{16}\}$, for the kernel of $D\Phi_{10}\vert_{D_{10}}$ whose elements each describe a one-parameter affine family stemming from $D_{10}$.  In Table~\ref{tab:16basis} we list the nonzero coordinates of each $\textbf{V}_1,\ldots,\textbf{V}_{16}$ -- each of which happens to be 1 or -1 -- by identifying the coordinates having either value. 

\begin{table}[htbp]
  \centering
  \caption{Nonzero coordinates of the vectors in the basis $\{\textbf{V}_1, \ldots, \textbf{V}_{16}\}$ for $D\Phi_{10}\vert_{D_{10}}$.  A nonzero coordinate has value 1 or -1, indicated by the subcolumn to which it belongs.}
  \renewcommand{\arraystretch}{.75}
  \setlength\tabcolsep{1.3mm}
  \begin{tabular}{crllllllllc|cllllllll}
       & & \multicolumn{8}{c}{\textbf{1}} & & & \multicolumn{8}{c}{\textbf{-1}} \\
      \cmidrule[1pt]{1-20}
      vector & & \multicolumn{18}{c}{coordinate}\\
      \cmidrule[1pt]{1-20}
     $\textbf{V}_1$ & & 2 & 3 & 7 & 8 & 74 & 75 & 79 & 80 & & & 10 & 18 & 19 & 27 & 55 & 63 & 64 & 72 \\
 $\textbf{V}_2$ & & 10 & 12 & 16 & 18 & 64 & 66 & 70 & 72 & & & 2 & 8 & 20 & 26 & 56 & 62 & 74 & 80 \\
    $\textbf{V}_3$ & & 28 & 29 & 35 & 36 & 46 & 47 & 53 & 54  & & &  4 & 6 & 13 & 15 & 67 & 69 & 76 & 78 \\
    $\textbf{V}_4$ & & 4 & 8 & 24 & 25 & 40 & 44 & 78 & 79  & & &  28 & 32 & 48 & 54 & 57 & 63 & 64 & 68 \\
    $\textbf{V}_5$ & & 37 & 40 & 47 & 54 & 55 & 58 & 65 & 72  & & &  5 & 7 & 15 & 17 & 32 & 34 & 78 & 80 \\
    $\textbf{V}_6$ & & 4 & 9 & 12 & 14 & 48 & 50 & 67 & 72  & & &   20 & 24 & 28 & 35 & 38 & 42 & 73 & 80\\
    $\textbf{V}_7$ & & 19 & 27 & 29 & 34 & 38 & 43 & 64 & 72  & & &   3 & 8 & 13 & 14 & 58 & 59 & 75 & 80 \\
    $\textbf{V}_8$ & & 37 & 39 & 43 & 45 & 46 & 48 & 52 & 54  & & &   5 & 6 & 23 & 24 & 59 & 60 & 77 & 78 \\
    $\textbf{V}_9$ & & 2 & 8 & 20 & 26 & 43 & 45 & 52 & 54  & & &   10 & 12 & 59 & 60 & 64 & 66 & 77 & 78 \\
    $\textbf{V}_{10}$ & & 47 & 48 & 49 & 50 & 74 & 75 & 76 & 77  & & &   15 & 18 & 24 & 27 & 33 & 36 & 42 & 45 \\
    $\textbf{V}_{11}$ & & 2 & 6 & 30 & 32 & 65 & 69 & 75 & 77  & & &   10 & 17 & 22 & 27 & 40 & 45 & 46 & 53 \\
    $\textbf{V}_{12}$ & & 12 & 14 & 30 & 32 & 48 & 50 & 75 & 77  & & &   20 & 22 & 24 & 27 & 38 & 40 & 42 & 45 \\ 
    $\textbf{V}_{13}$ & & 47 & 50 & 56 & 59 & 65 & 68 & 74 & 77  & & &   15 & 16 & 17 & 18 & 42 & 43 & 44 & 45 \\
    $\textbf{V}_{14}$ & & 25 & 26 & 34 & 35 & 47 & 50 & 74 & 77  & & &   15 & 18 & 42 & 45 & 57 & 58 & 66 & 67 \\
    $\textbf{V}_{15}$ & & 19 & 23 & 28 & 32 & 64 & 68 & 73 & 77  & & &   3 & 4 & 8 & 9 & 39 & 40 & 44 & 45 \\
    $\textbf{V}_{16}$ & & 19 & 23 & 49 & 53 & 58 & 62 & 73 & 77  & & &   3 & 9 & 33 & 34 & 39 & 45 & 69 & 70 \\
    \bottomrule
  \end{tabular}
  \label{tab:16basis}
\end{table}

For each $k=1,\ldots, 16$, the collection of matrices  $D_{10} \circ \text{EXP}\left( \ii \sigma_k \textbf{V}_k \right)$ -- where $\sigma_i$ is a parameter -- is an affine family.  This does not say that there is a 16-dimensional affine family stemming from $D_{10}$; in fact, simultaneous inclusion of all $\sigma_i$ destroys the Hadamard property for most choices of parameters.  However, many combinations of inclusion of parameters do give two- and three-dimensional families.  For example, 
$$
\setlength{\abovedisplayskip}{10pt}
\setlength{\belowdisplayskip}{10pt}
	D_{10}^{(3)}(a,b,c) = D_{10} \circ \text{EXP}\left( \ii (a\textbf{V}_1 + b\textbf{V}_2 + c\textbf{V}_3\right)).
$$
Another example of a three-parameter family found among these vectors is 
$$
\setlength{\abovedisplayskip}{10pt}
\setlength{\belowdisplayskip}{10pt}
	D_{10} \circ \text{EXP}\left( \ii (a\textbf{V}_4 + b\textbf{V}_{15} + c\textbf{V}_{16}\right)).
$$
Note that no combination of four (or more) of these vectors give an affine family, although this does not guarantee one does not exist. 

The existence of this particular basis for the kernel of $D\Phi_{10}\vert_{D_{10}}$ does provide one possibility for the genesis of a 16-dimensional center subspace:  certainly the defect must be at least 16 to account for the intersection of 16, possibly permutation-equivalent, one-parameter affine (sub-)families passing through $D_{10}$.

\section{Discussion}
\label{sec:discussion}
In this paper we have given a new interpretation of the defect of a Hadamard matrix as the dimension of a center subspace of a gradient flow whose fixed points are exactly the dephased Hadamard matrices. We have applied this technique to a simple example, using dynamical systems theory to explain why the defect of the real $4 \times 4$ Hadamard is larger than the complex members of the affine family $F_4^{(1)}$.  We have used tools from dynamical systems theory to prove that the $d \left(D_6(c)\right) = 4$ for all values of $c$ and have presented a new type of evidence in support of existing conjectures concerning $6 \times 6$ Hadamards.  Finally we have used our perspective to  uncover new $10 \times 10$ affine families.

It is a virtue of the formalism built in Section~\ref{sec:theory} that we need not have explicit values for the Taylor coefficients of the functions $\alpha_i(t_1,\ldots,t_c)$ to conclude that flow on the center manifold exists. It is a shortcoming of center manifold reduction, in general, that it cannot prove flow does \textit{not} exist on some part of a center manifold.  This limitation is a consequence of the fact that, \textit{a priori}, one has no knowledge of the smallest order in the Taylor expansion where one might first encounter nonzero contribution to the time-rates-of-change of the embedding parameters.  We are not certain that our application is bound with this deficiency and are hopeful that a deeper understanding of the derivatives of the vector field $\Phi_d$ may be exploited to further expand the use of dynamical systems theory to attack questions about complex Hadamards.  In particular, if one could show that all coefficients in the expansion of the time-rates-of-change of the embedding parameters vanish in the center manifold reduction of $\Phi_d$ at a Hadamard $H$, then one would prove the existence of a positive-dimensional family of complex Hadamards stemming from $H$.

Finally, vectors in a basis for the center subspace of $\Phi_n$ at an order-$n$ matrix $H$ may be tangent to affine families stemming from $H$.  This fact can be exploited to find new affine families, in the manner of Section~\ref{sec:examples}.

\end{document}